\documentclass[AMA,STIX1COL]{WileyNJD-v2}

\articletype{Article Type}%

\received{26 April 2016}
\revised{6 June 2016}
\accepted{6 June 2016}

\begin{document}

\title{A unifying modelling approach for hierarchical distributed lag models}

\author[1]{Theo Economou*}

\author[2]{Daphne Parliari}

\author[3]{Aurelio Tobias}

\author[4]{Laura Dawkins}

\author[5]{Oliver Stoner}

\author[4]{Hamish Steptoe}

\author[6,7,8]{Rachel Lowe}

\author[9]{Maria Athanasiadou}

\author[4]{Christophe Sarran}

\author[1,10]{Jos Lelieveld}

\authormark{T. Economou \textsc{et al.}}

\address[1]{\orgdiv{Climate and Atmosphere Research Centre - CARE-C}, \orgname{The Cyprus Institute}, \orgaddress{\state{Nicosia}, \country{Cyprus}}}

\address[2]{\orgdiv{Laboratory of Atmospheric Physics}, \orgname{Aristotle University}, \orgaddress{\state{Thessaloniki}, \country{Greece}}}

\address[3]{\orgdiv{Institute of Environmental Assessment and Water Research}, \orgname{Spanish Council for Scientific Research}, \orgaddress{\state{Barcelona}, \country{Spain}}}

\address[4]{\orgdiv{Met Office}, \orgaddress{\state{Exeter}, \country{UK}}}

\address[5]{\orgdiv{School of Mathematics and Statistics}, \orgname{University of Glasgow}, \orgaddress{\state{Glasgow}, \country{UK}}}

\address[6]{\orgdiv{Barcelona Supercomputing Center (BSC)}, \orgaddress{\state{Barcelona}, \country{Spain}}}

\address[7]{\orgdiv{Catalan Institution for Research and Advanced Studies (ICREA)}, \orgaddress{\state{Barcelona}, \country{Spain}}}

\address[8]{\orgdiv{London School of Hygiene \& Tropical Medicine}, \orgaddress{\state{London}, \country{UK}}}

\address[9]{\orgdiv{Health Monitoring Unit}, \orgaddress{\state{Ministry of Health}, \country{Cyprus}}}

\address[10]{\orgdiv{Max Planck Institute for Chemistry}, \orgaddress{\state{Mainz}, \country{Germany}}}

\corres{*Theo Economou. \email{t.economou@cyi.ac.cy}}

\presentaddress{This is sample for present address text this is sample for present address text}

\abstract[Summary]{We present a statistical modelling framework for implementing Distributed Lag Models (DLMs), encompassing several extensions of the approach to capture the temporally distributed effect from covariates via regression. We place DLMs in the context of penalised Generalized Additive Models (GAMs) and illustrate that implementation via the R package \texttt{mgcv}, which allows for flexible and interpretable inference in addition to thorough model assessment. We show how the interpretation of penalised splines as random quantities enables approximate Bayesian inference and hierarchical structures in the same practical setting. We focus on epidemiological studies and demonstrate the approach with application to mortality data from Cyprus and Greece. For the Cyprus case study, we investigate for the first time, the joint lagged effects from both temperature and humidity on mortality risk with the unexpected result that humidity severely increases risk during cold rather than hot conditions. Another novel application is the use of the proposed framework for hierarchical pooling, to estimate district-specific covariate-lag risk on morality and the use of posterior simulation to compare risk across districts.}


\jnlcitation{\cname{%
\author{T. Economou},
\author{D. Parliari},
\author{A. Tobias},
\author{L. Dawkins},
\author{O. Stoner}, and
\author{J. Lelieveld}} (\cyear{2023}),
\ctitle{Hierarchical DLNMs}, \cjournal{Statistics in Medicine}, \cvol{2017;00:1--6}.}

\maketitle

\footnotetext{\textbf{Abbreviations:} DLM: Distributed Lag Model; DNLM: Distributed Non-Linear Model; GAM: Generalized Additive Model; RR: Relative Risk; CR: Cumulative Risk.}

\section{Introduction}\label{sec:intro}
Distributed lag models (DLMs) constitute an extremely useful statistical modelling class for characterising the effects of a covariate and its temporally lagged values on the mean of a response variable. First introduced in econometrics~\cite{Almon1965}, they have now become the de facto modelling tool in epidemiological studies (e.g., \citep{Gosia2015,humidDLNM2019,Roye2020,Tsangari2016}), when interest lies in linking environmental covariates with health outcomes (e.g., linking heat and mortality). %The framework here is presented with an epidemiological focus and in the context of count data, although the approach is applicable to any type of data. 
For some count variable $y_t$ (e.g., mortality counts) and some environmental covariate $x_t$ (e.g., temperature) a DLM can be generically defined as
\begin{eqnarray}
	y_t &\sim& Poisson(\mu_t) \label{eq:NegBin}\\
	\log(\mu_t) &=& \alpha + \beta_0x_t + \beta_1x_{t-1} + \cdots + \beta_{L}x_{t-L}.\label{eq:DLM}
\end{eqnarray}
The mean count is $\mu_t$ and $L$ is the maximum number of lags. This is a Generalized Linear Model (GLM), albeit one that is likely to be unstable due to high correlation across the lagged versions of $x_t$. One remedy is to constrain the coefficients $\beta_l$, e.g. by assuming they come from some unknown smooth function $g(\cdot)$ such that $\beta_l = g(l)$. Other ways of constraining the coefficients are possible~\cite{Armstrong2006,Richardson2009}, including Bayesian methods that assume the coefficients are random~\cite{Obermeier2015} or have priors that reflect scientific understanding~\cite{Welty2009}. Here we argue that using smooth but penalised functions is an objective approach to constraining the $\beta_l$'s, in addition to providing some robustness to the choice of $L$.%, resulting in low reliance on subjective expert information to inform the model design.

DLMs have been extended to distributed lag \emph{non-linear} models (DLNMs) so that the effect of the covariate at each lag is non-linear~\cite{Armstrong2006,Gasparini2010}. This greatly increases modelling flexibility so that the linear predictor in~\eqref{eq:DLM} becomes
\begin{equation}\label{eq:DLNM}
	\log(\mu_t) = \alpha + h(0,x_t) + h(1,x_{t-1}) + \cdots  + h(L,x_{t-L})	
\end{equation}
where $h(l,x_{t-l})$ is a two dimensional function of lag and the covariate. A further important extension to DLNMs is in the interaction of the covariate-lag relationship with other factors such as other environmental variables or indeed other dimensions such as space, season etc. For example,~\cite{Muggeo2007} propose a model with an interaction between temperature, lag and air-quality, whereas~\cite{Obermeier2015} fit a model that interacts the covariate-lag function with a spatial Markov random field to allow for spatial structure in the relationship. Moreover, a two-stage approach~\cite{Dominici2000,Gasparrini2012,Gasparrini2013} has been proposed to pool estimates from individual DLNM fits across spatial regions using a meta-analysis modelling framework. 

In this paper we introduce a modelling approach for fitting DLNMs that encompasses all aforementioned extensions into a unified framework. Specifically we propose to fit hierarchical DLNMs using the machinery of penalised Generalised Additive Models (GAMs) as implemented in the R package \texttt{mgcv}~\cite{Wood2011,Wood2017}. The hierarchical nature of the approach is particularly important when considering covariate interactions since hierarchical structures are a natural way of dealing with sparse data.% and the `curse of dimensionality'~\cite{Gelman2013} -- the fact that computationally, the methods may not scale well with higher dimensions and will require a enormous amount of data for reliable estimation. 
GAMs from the \texttt{mgcv} package have been used before to fit DLNMs across a range of application areas (e.g., dendrochronology~\cite{Nothdurft2018} and earthquakes~\cite{Obermeier2015}) and one of the main contributions here is to illustrate further the use of GAMs for flexible modelling involving the effect from lagged covariates, including hierarchical structures across discrete categories (regions, age-groups etc.). We present the framework in the context of epidemiological data and place particular emphasis on scientific utility by focusing on a) uncertainty quantification through Bayesian simulation-based inference, b) thorough model checking, c) interpretability of the results and d) computational efficiency. 

In section~\ref{sec:framework} we present the framework and describe in detail how DLNMs can be interpreted as penalised GAMs. We discuss fitting, inference, model checking and the inclusion of hierarchies. Then in section~\ref{sec:implementation} we demonstrate their application with real data, specifically mortality data from the city of Thessaloniki, Greece and the island of Cyprus. Finally, section~\ref{sec:conclusion} presents a summary, some conclusions and discusses caveats and future directions.

\section{DLNMs as Hierarchical GAMs}\label{sec:framework}
In a GAM based on penalised regression splines, the effect of covariate $x$ can be represented by an unknown smooth function:
\begin{equation}
f(x_i) = \sum_{k=1}^K \beta_k b_k(x_i) = \boldsymbol{X}_i\boldsymbol{\beta} \label{eq:basis}
\end{equation}
where $b_k$ are basis functions, $\beta_k$ are unknown coefficients and $\boldsymbol{X}$ the resulting model matrix. The smooth function $f(\cdot)$ is penalised when $K$ (the number of coefficients) is too large, which would result in $f(\cdot)$ being too "wiggly" and thus the model over-fitting the data. Inference is based on penalised log-likelihood via:
\begin{equation}\label{eq:penLik}
	\ell(\boldsymbol{\beta},\phi;\boldsymbol{y}) - \lambda\int_{x}f''(x)^2 dx= \ell(\boldsymbol{\beta},\phi;\boldsymbol{y}) - \lambda\boldsymbol{\beta}'\boldsymbol{S}\boldsymbol{\beta}
\end{equation}
where $\ell(\cdot)$ is the log-likelihood, $\lambda$ is a penalty parameter and $\boldsymbol{S}$ is a penalty matrix that relates to a quadratic penalty on $\boldsymbol{\beta}$. Matrix $\boldsymbol{S}$ depends on the choice of basis functions as well as any constraints on the function (such as being centered on zero to identify an overall intercept term). In practice, $K$ is chosen to be larger than necessary while penalisation via $\lambda$ guards against over-fitting. The amount of penalisation is estimated from the data as a compromise between out-of-sample and in-sample predictive skill~\cite{Wood2017}.%, e.g. using quantities that approximate out-of-sample prediction error such as the AIC (Akaike Information Criterion) or the GCV (Generalised Cross Validation).

\subsection{Bayesian GAMs}
The smoothness of $f(\cdot)$ in \eqref{eq:basis} can be viewed as a constraint on the values of $\boldsymbol{\beta}$ and from a Bayesian viewpoint, this can be represented with a prior distribution $\boldsymbol{\beta} \sim N(\boldsymbol{0}, \boldsymbol{S}^{-}/\lambda)$, where $\boldsymbol{S}^{-}$ is the pseudo-inverse of $\boldsymbol{S}$. Assuming the coefficients are random, allows estimation by restricted maximum likelihood (REML). Conditional on estimates of the penalty parameter and any hyperparameters $\phi$, the posterior distribution can then be approximated~\cite{Wood2017} by:
\begin{equation}\label{eq:posterior}
	\boldsymbol{\beta}|\boldsymbol{y},\lambda,\phi \sim N\left(\hat{\boldsymbol{\beta}}, (\boldsymbol{X}'\boldsymbol{W}\boldsymbol{X}+\hat{\lambda}\boldsymbol{S})^{-1}\hat{\phi}\right),
\end{equation}
where $\boldsymbol{W}$ is the weight matrix associated with the penalised iterative re-weighted least squares algorithm used to fit GAMs~\cite{Wood2017} and $\hat{\boldsymbol{\beta}}$, $\hat{\phi}$, $\hat{\lambda}$ are the REML estimates. All components of~\eqref{eq:posterior} are readily provided when fitting GAMs in \texttt{mgcv}.

The posterior predictive distribution (PPD) of any value of the response $\tilde{y}$ given the data can then be obtained by
\begin{equation}\label{eq:post_pred}
	p(\tilde{y}|\boldsymbol{y},\lambda,\phi) = \int_{\boldsymbol{\beta}} p(\tilde{y}|\boldsymbol{\beta},\lambda,\phi) p (\boldsymbol{\beta}|\boldsymbol{y},\lambda,\phi) d\boldsymbol{\beta}.
\end{equation}
Note the compromise in that not all estimation (sampling) uncertainty is being integrated out, since posterior distributions for $\lambda$ and $\phi$ are not readily available. The predictive distribution is defined conditionally on the penalty parameter(s) $\lambda$ and any hyperparameters $\phi$. We consider this compromise to be acceptable given the computational efficiency of using this approach (e.g., compared to a full MCMC fit as presented in the supplementary material where quantification of this compromise is also given for a particular case study). In practice, Monte Carlo simulation is used to approximate~\eqref{eq:post_pred} by simulating from~\eqref{eq:posterior} and then from the distribution of the response $p(y|\boldsymbol{\beta},\lambda,\phi)$.

\subsection{Interactions}
Smooth functions of multiple covariates can be constructed using tensor product smooths. A function of $x$ and $z$ say, can be defined by assuming that the coefficients of~\eqref{eq:basis} are smooth functions of $z$ e.g.,
\begin{eqnarray}\label{eq:tensor}
	f(x_i,z_i) &=& \sum_{k=1}^K \beta_k(z_i) b_k(x_i) \quad \mbox{ where } \quad \beta_k(z_i) = \sum_{j=1}^J \gamma_{k,j} a_j(z_i) \\
	\implies f(x_i,z_i) &=& \sum_{k=1}^K \sum_{j=1}^J \gamma_{k,j} a_j(z_i)b_k(x_i) = \boldsymbol{P}_i\boldsymbol{\gamma}
\end{eqnarray}
where $\gamma_{k,j}$ are coefficients, $a(\cdot)$ are basis functions and $\boldsymbol{P}$ is the resulting ($n\times (J\cdot K)$) model matrix. Higher dimensional functions can be constructed in the same way at the expense of exponential growth in the number of coefficients.

\subsection{DLNMs as GAMs}
DLNMs can therefore be seen as GAMs, where $h(l,x_{t-l})$ in~\eqref{eq:DLNM} is constructed as a tensor product and the model fitted using the \texttt{linear.functional.terms} option in \texttt{mgcv}. It is now straightforward to consider lagged effects of more than one covariate (e.g., temperature $x_t$ and humidity $z_t$) by extending equation~\eqref{eq:DLNM} to include terms such as $h(l,x_t,z_t)$. This greatly increases the current functionality of DLNMs to investigate the lagged effect of a covariate as a function of another, demonstrated in section~\ref{sec:implementation}.

Smoothness and penalisation of the tensor products ensures optimal use of the data. Smoothing is a form of data pooling since, for example, the temperature-lag relationship at neighbouring humidity values will be similar. Penalisation ensures that the coefficients are constrained so that even if a large number of them is considered to begin with, over-fitting is avoided. This is a desirable property since multidimensional functions require a lot of data to be estimated reliably.%, a luxury not always afforded when working with health data.

\subsection{Hierarchies}\label{sec:hierarchies}
Smoothing, however, only works with numerical (non-categorical) covariates. For instance, we may want to consider a separate temperature-lag relationship for different spatial regions, albeit with some pooling across the regions. With tensor product interactions, we can model a covariate-lag relationship that varies smoothly in space with a term such as $h(l,x_{t-l},z_{1,s},z_{2,s})$, where $z_{1,s}$ and $z_{2,s}$ are spatial coordinates of the centroid of some region $s$. Spatial smoothness in $h(l,x_{t-l},z_{1,s},z_{2,s})$ will force regions in close proximity to have a similar temperature-lag effect. When this is not desirable (e.g., when regions are heterogeneous due to region-specific characteristics such as socioeconomic factors, even when they are in close proximity) then this formulation is not appropriate. A different way of achieving the desired pooling is to assume the categorical units (regions) are exchangeable, and construct a hierarchical structure with a "global" term plus unit-specific deviations. 

This is a hierarchical GAM~\cite{Pedersen2019}: 
\begin{equation}
	\log(\mu_{t,s}) = \alpha + \alpha_s + \sum_{l=0}^L h(l,x_{t-l}) + \sum_{l=0}^L h_s(l,x_{t-l}) = \alpha + \alpha_s + f(x_t) + f_s(x_t) \label{eq:DLNM_hier} 
\end{equation}
where $f(x_t)$ is interpreted as the global term (the overall mean temperature-lag effect across all units $s$), and $f_s(x_t)$ are the unit-specific deviations from the global. As such, $(f(x_t) + f_s(x_t))$ are the individual district-specific effects. The presence of a global term induces the pooling. If there is little difference across units $s$, the functions $f_s(x_t)$ will be much simpler (in terms of number of parameters) than $f(x_t)$ due to penalisation, and therefore a more optimal use of the data is achieved. The intercept $\alpha$ is the global mean (at the log scale), while $\alpha+\alpha_s$ are the unit-specific intercepts, and where $\alpha_s\sim N(0,\sigma^2)$ is a random effect. The $\alpha_s$ effects are also constructed via regression splines (using a ridge penalty) in \texttt{mgcv}.

\subsubsection{Amount of pooling}
In GAMs, one can force each deviation to have the same amount of smoothness assuming a shared penalty parameter. Forcing the deviations to have the same smoothness induces pooling, in much the same way that i.i.d. Gaussian random effects share the same variance (e.g., $\beta_j = \beta+\theta_j$ where $\theta_j\sim N(0,\sigma^2)$). To achieve this in \texttt{mgcv}, the ``deviations'' $h_s(l,x_{t-l})$ can be constructed as tensor products of lag $l$, covariate(s) $x_{t-l}$ and a marginal basis for $s$ that emulates the behaviour of an i.i.d. Gaussian random effect (in the same way as $\alpha_s$ in equation~\eqref{eq:DLNM_hier}). In \texttt{mgcv}, this is done by exploiting the duality between Gaussian random effects and splines (\cite{Wood2017}, Ch. 5), that allows the covariate-lag effect to be interacted with an i.i.d. random effect over $s$ resulting in shrinkage of $\left(f(x_t) + f_s(x_t)\right)$ towards $f(x_t)$. 

The amount of pooling can be controlled via the number of coefficients in each tensor product. For instance, allowing the global term $h(l,x_{t-l})$ a much smaller number of coefficients compared to the deviations $h_s(l,x_{t-l})$, will result in much less pooling than vice versa. This increases the amount of subjectivity in the estimation (which may be desirable if, for instance, the estimates turn out to be counter-intuitive relative to scientific understanding). Pooling in hierarchical models aims to provide as robust estimates as possible given the data, although admittedly it is simply a statistical ``trick'' to allow the inference to borrow information across the data structures. For instance, the global term $h(l,x_{t-l})$ is not necessarily an interpretable quantity (given the penalisation that takes place), it is there to ensure pooling. If the goal is a global estimate, then a global model would be more appropriate.

\subsection{Interpretation and inference}\label{sec:inferencce}
%All models presented are implemented in the R package \texttt{mgcv}, using REML, as this enables the interpretation of the splines as random quantities. This in turn enables Bayesian inference and prediction via Monte Carlo simulation, of all associated quantities since the covariance matrix of the approximate multivariate Gaussian posterior (equation \eqref{eq:posterior}) of the splines coefficients is readily provided by \texttt{mgcv}. Simulating from this posterior then enables the full quantification of associated quantities.

In the context of modelling counts of health outcomes, the (log) mean count for a single covariate $x_t$ can be modelled as:
\begin{equation}
	\log(\mu_t) = \alpha + \sum_{l=0}^L h(l,x_{t-l}) + \log(O_t).
\end{equation}
The splines are by default constrained to be centered at zero, so that $h(l,x_{t-l})$ is the additive change in the overall log mean count or, in the presence of an offset $O_t$, it is the change in the mean occurrence rate. For instance if $O_t$ relates to the number of people at risk, then $h(l,x_{t-l})$ is the change in the log-mean mortality rate per unit population. Equivalently, $RR(l,x_{t-l})=\exp\{h(l,x_{t-l})\}$ can be interpreted as the relative risk (RR), the multiplicative change with respect to $\exp\{\alpha\}$, the mean count or rate across the time period of study. Sometimes it is desirable to compute the RR in relation to mortality at a specific value of the covariate say $\tilde{x}$, which can be computed by $RR(l,x_{t-l}) = \exp\{h(l,x_{t-l}) - h(l,\tilde{x})\}$.

The conventional way of illustrating the estimated effects from DLNMs is a plot of $RR(l,x)$ over a grid of finite values for $x$ and $l$. This is a 3D plot (e.g., Figure~\ref{fig:figure1}a) showing how risk varies for values of $x$ across different lags (note, one can alternatively display the same information as 2D ``heat maps'', e.g., Figure~\ref{fig:figure2}, which presents the whole surface more clearly). The plot is a counterfactual ``statement": keeping $x_{t-l}$ fixed for all lags $l$ (e.g., fixing temperature at 40$^\circ$C for $L=20$ days) and seeing what the associated risk contributions are from each lag. A useful summary of the effects in the covariate dimension is the cumulative risk (CR) where the additive effects $h(l,x_{t-l})$ are summed across the lags. I.e.,
\begin{equation}\label{eq:CR_x}
	CR(x_t) = \exp\left\{ \sum_{l=0}^L h(l,x_{t-l}) \right\}
\end{equation}
which quantifies the total risk under the counterfactual assumption of $x_t$ being the same for all $l$. %Equivalently we can compute 
%\begin{equation}\label{eq:CR_l}
%	CR(l) = \exp\left\{ \sum_{x=x_{low}}^{x_{high}} h(l,x) \right\}
%\end{equation}
%which quantifies the total risk at each lag, across a (discrete) range of covariate values $[x_{low},x_{high}]$.

Since $h$ is a function of the spline coefficients, any inference on quantities that are functions of $h$ (such as RR and CR) can be performed using Monte Carlo simulation. For instance, a 95\% (posterior) credible interval can be constructed for $RR(l,x)=\exp\{h(l,x)\}$ to check whether the value 1 lies inside this interval. This would indicate strong weight of evidence for the risk not being significantly different from the baseline risk. 

Furthermore, posterior predictive simulation of the counts $y_t$ (equation \eqref{eq:post_pred}) enables thorough model checking (\cite{Gelman2013}, Ch. 6). Note however that for computational efficiency, the Bayesian simulation-based inference that is presented here ignores estimation uncertainty in the penalty parameters and any hyperparameters (of the conditional distribution). However, full Bayesian inference is possible for all models presented here, via MCMC through the R package \texttt{nimble}~\cite{nimble} and the \texttt{mgcv} function \texttt{jagam}. A full working example is provided in the online supplementary material.

In the following section, we present application of the various modelling options to real data and also discuss more interpretable risk quantities from the models that are not based on counterfactual events. %In addition, we perform a simulation experiment to explore operational characteristics and assess the performance of the hierarchical models. 

\section{Implementation and case studies}\label{sec:implementation}
All models presented here are implemented in the R package \texttt{mgcv}, using REML, as this enables the interpretation of the splines as random quantities. This section presents the implementation of the framework to the typical scenario of linking environmental covariates (exposures) to human mortality. Firstly, we present a straightforward application of a DLNM with a single covariate and present a comparison with conventional implementation approaches, interpretation of results and model checking. Then, we present the case where two or more covariates are of interest, including a situation with hierarchical (spatial) structure.

\subsection{Single lagged covariate}\label{sec:Thess}
This example involves the effect of maximum daily apparent temperature or Tapp (a heat stress indicator that is a function of temperature and relative humidity) on human mortality in the city of Thessaloniki, Greece. The data have been studied in~\cite{Parliari2022} and consist of daily all-cause death counts (excluding accidents) for 2006-2016. Interest lies in the effect on mortality from Tapp or $x_t$, and its lags of up to $L=20$ days. We consider the following (baseline) model:
\begin{eqnarray}
    y_t &\sim& Poisson(\mu_t) \label{eq:ThesModel1} \\
	\log(\mu_t) &=& \alpha + \sum_{l=0}^L h(l,x_{t-l}) \label{eq:ThesModel2}
\end{eqnarray}
where $h$ is a tensor product of two marginal thin-plate regression splines (TPRS), each with 10 knots. The TPRS basis is a particularly attractive choice as it avoids explicit knot placement~\cite{Wood2017}. The function has $10^2-1=99$ coefficients (the minus one is for the center-on-zero constraint). The function \texttt{k.check()} in \texttt{mgcv} provides an assessment of whether the number of coefficients is adequate and here, it indicates that after penalisation the effective degrees of freedom (edf) is only $\approx$27 (out of 99), suggesting that 99 coefficients are more than enough. 

\begin{figure}[h]
	\centerline{
	\includegraphics[width=0.3\textwidth]{"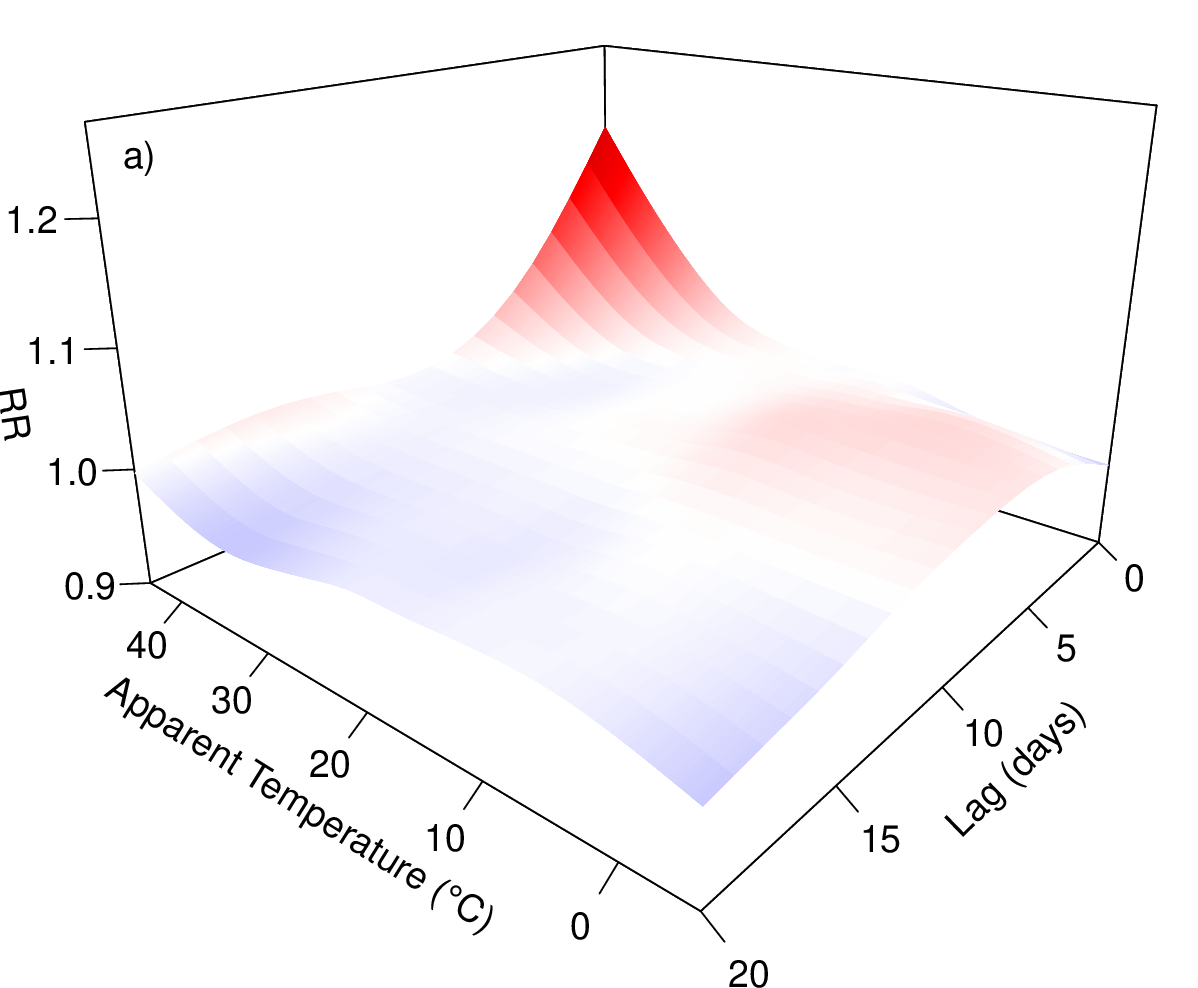"}
	\includegraphics[width=0.3\textwidth]{"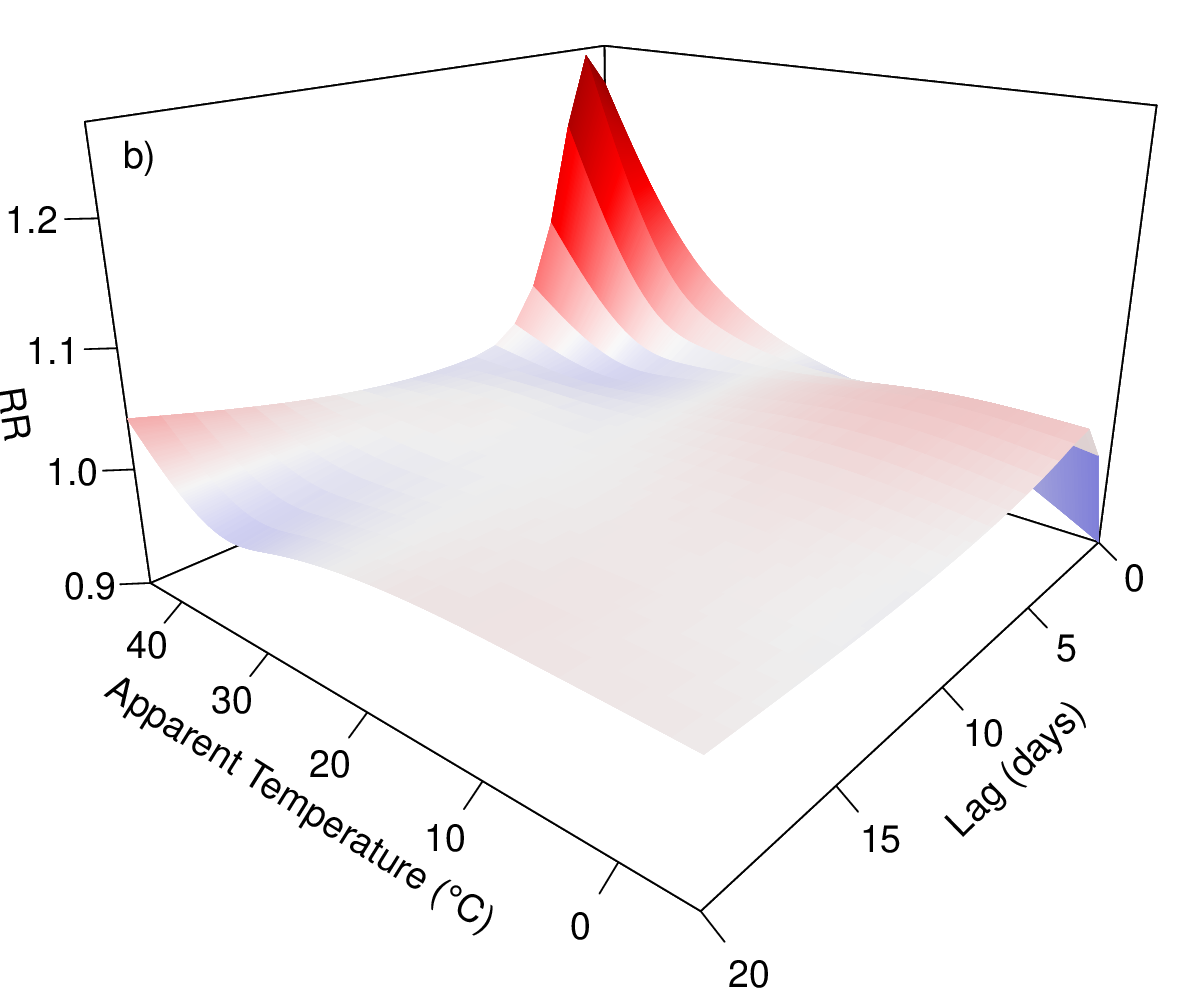"}
	\includegraphics[width=0.3\textwidth]{"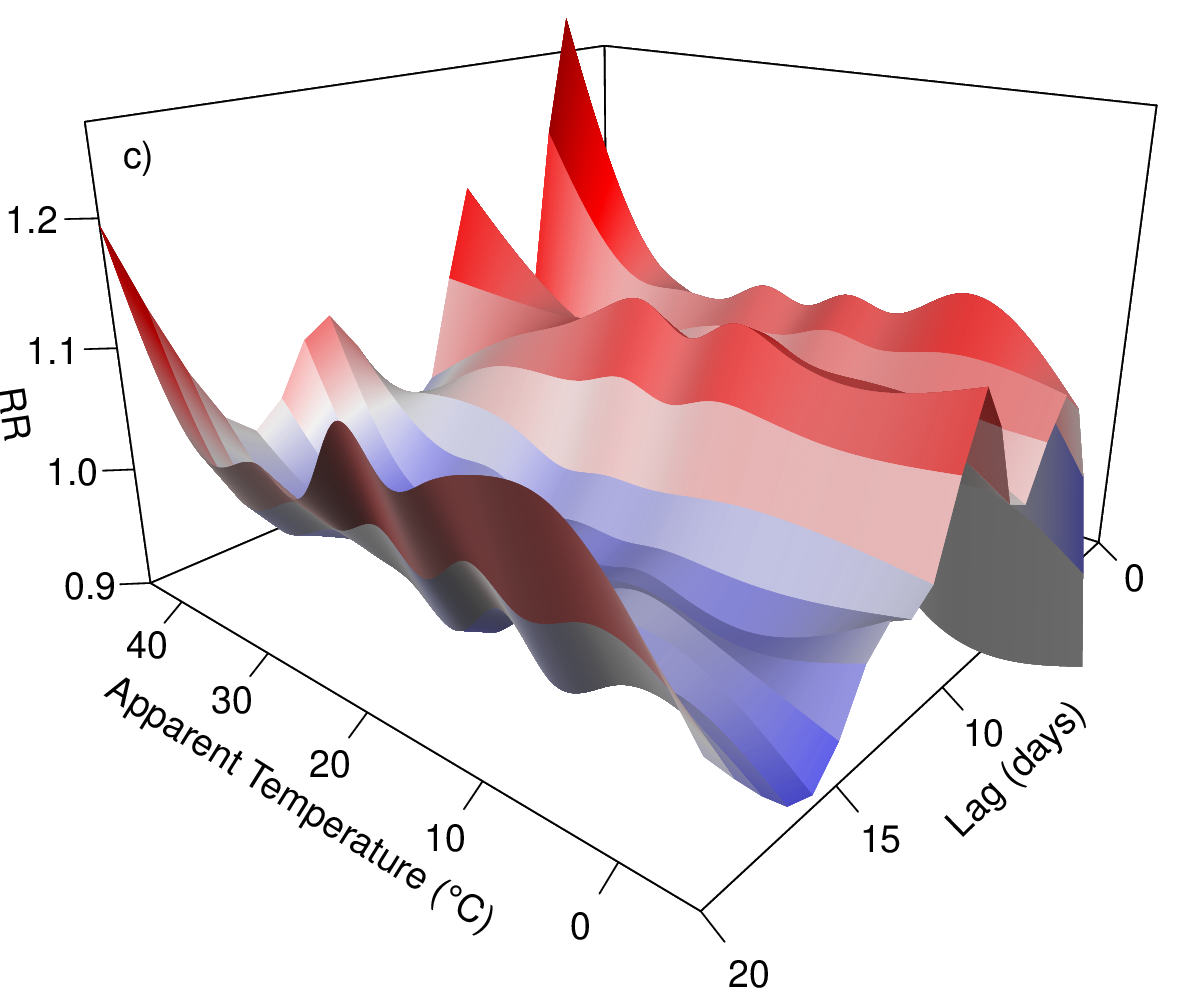"}
	\includegraphics[height=4.2cm]{"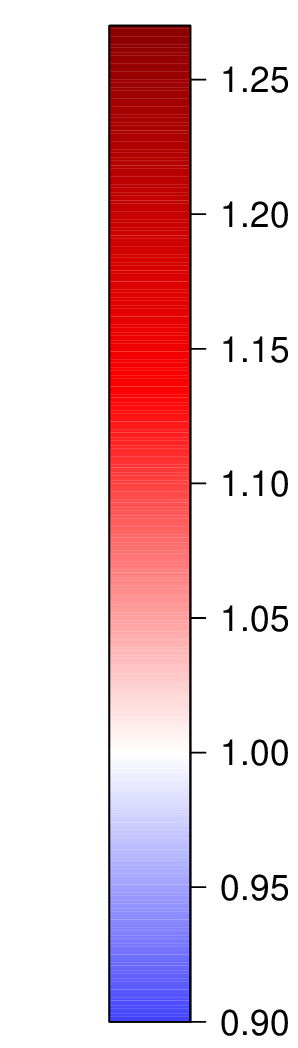"}
	}
	\caption{Relative risk estimates for the Thessaloniki data. Red colour indicates RR values $>1$ while blue colour signifies RR values $<1$. Panel a) shows RR estimates from a Poisson GAM, panel b) shows estimates from the original quasi-Poisson model fitted to this data and panel c) shows estimates from a Poisson GAM but without penalisation.
		\label{fig:figure1} }
\end{figure}

The estimated $RR(l,x)$ is shown in Figure~\ref{fig:figure1}a, for Tapp values spanning the data range. Function $h$ is centered at zero so the risk is relative to the overall mean mortality count, estimated to be $\exp\{\hat{\alpha}\}$=18.43 (cf. sample mean $\bar{y}=18.42$). There is elevated risk ($\approx$20\% above average) for very high Tapp values for up to 5 lags (days), while the rest of the surface appears flat around $RR=1$. Figure~\ref{fig:figure1}b shows the same estimate based on a DNLM fitted using the \texttt{dlnm} package~\cite{dlnm_package}, emulating the original analysis of this data in~\cite{Parliari2022}. This model is also Poisson, but $h$ was constructed as an interaction of two natural splines, with 4 and 5 coefficients for $x$ and $l$ respectively. There were therefore 20 coefficients and the fit did not involve penalisation. The surfaces in Figures~\ref{fig:figure1}a and~\ref{fig:figure1}b are qualitatively the same, but the extremes are less pronounced in the GAM (despite it having more coefficients). Penalisation is therefore important in performing inference in DLNMs as the risk estimates can be very sensitive to the number of coefficients (see also~\cite{Gasparrini2017}). For instance, fitting model~\eqref{eq:ThesModel1}--\eqref{eq:ThesModel2} without penalisation, results in the estimated surface shown in Figure~\ref{fig:figure1}c. The peak around $40^\circ$C and lag 20 is in contrast to current scientific evidence and probably ``noise'' rather than ``signal''. 

\begin{figure}[h]
	\centerline{
		\includegraphics[width=0.5\textwidth]{"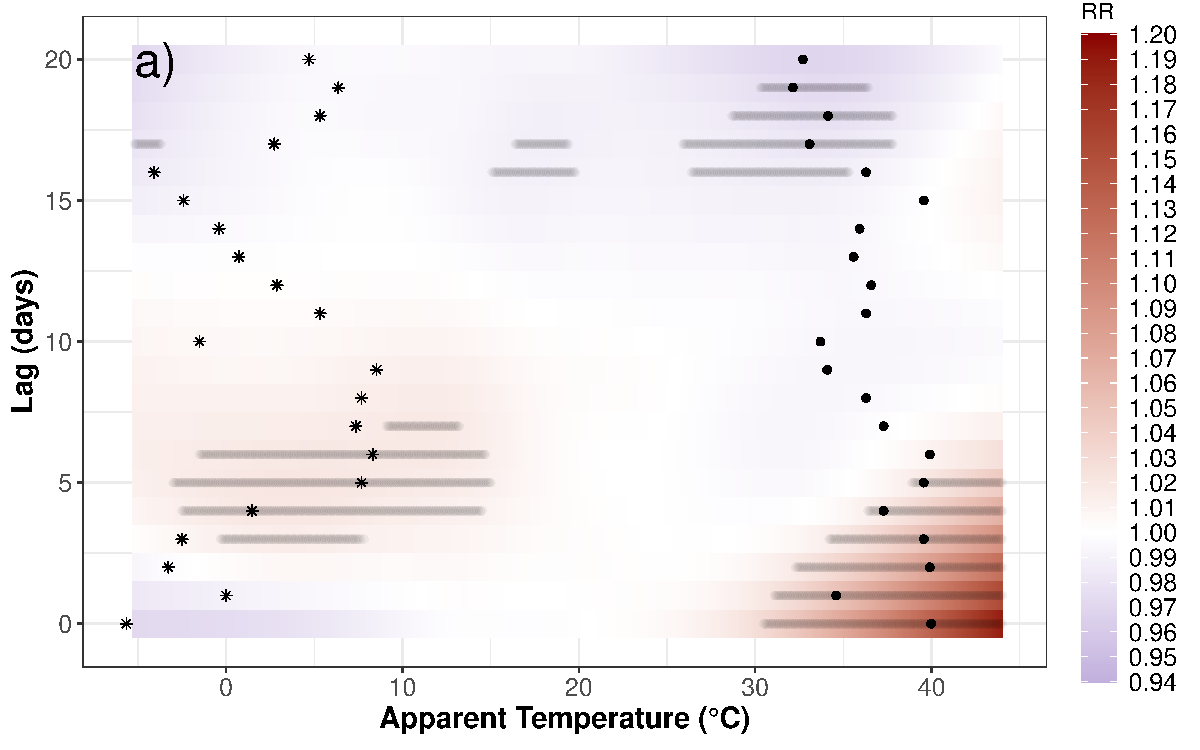"}
		\includegraphics[width=0.5\textwidth]{"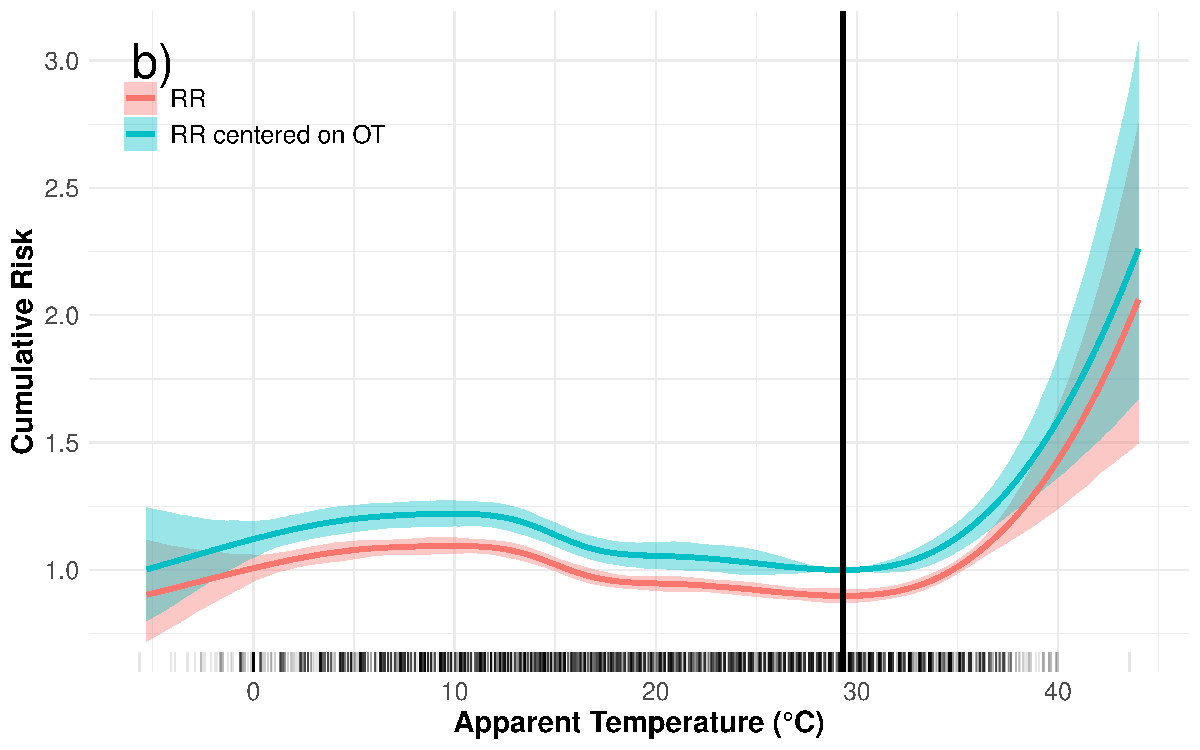"}
	}
\caption{Risk estimates for the Thessaloniki data. Panel a) shows RR estimates with regions of significant difference from RR $=1$ (at 95\% level) shaded in grey. The points show the temporal trajectory of a hot day (24/Jul/2007) while stars indicate the trajectory of a cold day (09/Feb/2006). Panel b) shows the resulting cumulative risk in red, and also in blue for when the baseline risk reflects the OT (vertical line).  
    \label{fig:figure2} }
\end{figure}

The weight of evidence in different regions of the temperature-lag range space is shown (in grey) in Figure~\ref{fig:figure2}a, which is the same as Figure~\ref{fig:figure1}a but in 2D. The elevated risk over short lags and high Tapp (bottom right corner) is significant. Another two ``patches'' of significance are evident. One over moderate lags and Tapp $\in [-3,15]$, an effect attributed in the literature to elevated transmission of communicable diseases such as influenza during cold temperatures~\cite{Armstrong2019,Barreca2012,Zeng2017}. The other patch of lower-than-average RR region at long lags and Tapp $\in [25,35]$ reflects the ``harvesting paradox'', where vulnerable population have perished during intense heat over short lags while the surviving ``more resilient'' population appears as lower-risk.

To investigate the marginal effect of the covariate (Tapp), the red curve in Figure~\ref{fig:figure2}b shows the associated cumulative risk $CR(x)$ (equation~\eqref{eq:CR_x}). Note the elevated risk above 35$^\circ$C but also in the region 5--15$^\circ$C. %The increased uncertainty in two extremes reflects the amount of data (indicated by the ``rug'' at the bottom of the plot). 
The Tapp value corresponding to the minimum $CR(x)$ has been termed as the ``optimum temperature'' or OT in~\cite{Honda2014}, here approximated at 29.3$^\circ$C. Changing the baseline risk to be the mortality risk at this Tapp value (section~\ref{sec:inferencce}), yields the blue curve in Figure~\ref{fig:figure2}b.

\subsection{Attributable fraction}
Both the $RR$ and the $CR$ are counterfactual quantities as they do not allow for the variability in the distribution of the covariate. For instance, a vertical ``slice'' along the lag dimension in Figure~\ref{fig:figure2}a assumes that Tapp is fixed to the same value for 21 days (the likelihood of which is probably very small). Contrast this with some actual Tapp trajectories shown in Figure~\ref{fig:figure2}a for a very hot and a very cold day.

For this reason, metrics such as the Attributable Fraction (AF) have been proposed~\cite{Steenland2006}, to interpret the risk estimates in terms of the observed data. If $R_0$ is the risk when Tapp is equal to the OT, then it can be interpreted as the risk of the ``least-exposed'' population. Then if $R_1$ is the risk due to any other covariate value ($R_1>R_0$ by definition of the OT), we can define $AF=1-(R_0/R_1)$ as the fraction of mortality cases attributable to the covariate being different to the OT. For DLNMs, we use CR to define these two risks, so for \emph{observed covariate} $x_t$ on day $t$
\begin{equation}\label{eq:AF}
AF(x_t) = 1 - \frac{\exp\left\{ \sum_{l=0}^L h(l,OT) \right\}}{\exp\left\{ \sum_{l=0}^L h(l,x_{t}) \right\}} = 1 - \exp\left\{ \sum_{l=0}^L \left[h(l,x_{t}) - h(l,OT)\right] \right\}.
\end{equation}
Note the use of of $x_t$ rather than $x_{t-l}$ reflecting how the risk of experiencing $x_t$ ``today'' is distributed over the next $L$ days. This was termed as the forward AF in~\cite{Gasparrini2014}. %, who provide a thorough exposition of attributable risk in DLNMs. The backward AF case (that uses the observed sequences of $x_{t-l}$ for $l=0,\ldots,L$) is not discussed here, although the code for computing it is provided. As shown in section~\ref{sec:multiple_exposures}, using the forward AF is more natural to use when considering more than one covariate.
Then, the attributable number $AN(x_t)=AF(x_t)\cdot y_t$ is the number of cases that are attributable to $x_t$ on day $t$, where $y_t$ is the observed mortality count. Finally, an overall estimate of the $AF$ is obtained via $\bar{AF}=\sum_tAN(x_t)/\sum_t y_t$ and here this was estimated as $\bar{AF}=0.100$ with 95\% credible interval [0.07, 0.125], meaning that about 10\% of deaths in Thessaloniki can be attributed to sub-optimal Tapp values.

It is also instructive to quantify the AF for extremely high/low Tapp, termed heat-related AF or cold-related AF. The estimated AF for $\mbox{Tapp}>35^\circ$C (the 0.95 sample quantile) was 0.196 [0.142, 0.247], while for $\mbox{Tapp}<2.6^\circ$C (the 0.05 sample quantile) it was 0.111 [0.05, 0.167]. The heat-related AF is then about twice as large for Thessaloniki. Computing the heat-related or cold-related AF is more straightforward using the forward AF, which is another reason it is preferred here.

\subsection{Model Checking, Selection and Expansion}
Posterior predictive model checking~\cite{Gelman2013} compares observations of the response (mortality counts) against their respective predictive distribution according to the model. This has the advantage of naturally encompassing model selection and expansion.

\subsubsection{Overdispersion}
A well-known constraint of Poisson regression models is that the mean equals the variance, usually resulting in underestimation of the variability in the data. Traditionally (and under asymptotic assumptions) a hypothesis test involving the deviance of the fitted model~\cite{Wood2017} can be used to check for overdispersion. For the Thessaloniki Poisson model the p-value from such a test is effectively zero indicating overdisperion is an issue.

The usual remedy to overdisperion is to use quasi-Poisson or the Negative Binomial distribution. Both options introduce a dispersion parameter to ``inflate'' the mean accordingly to avoid overdisperion. In terms of estimates (coefficients, relative risk etc.) and their uncertainty, both approaches yield virtually identical results (see Figure S1 and associated code in  online supplementary material). However, with quasi-Poisson there is no analytical expression for the distribution prohibiting posterior predictive checking, so here we opt for the Negative Binomial distribution. We thus fit model~\eqref{eq:ThesModel1}--\eqref{eq:ThesModel2} again but with $NegBin(\mu_t,\theta)$ rather than $Poisson(\mu_t)$, where $\theta$ is the dispersion parameter. This guarantees a favourable p-value of the hypothesis testing above, however a more thorough check can be done using posterior predictive checking.  

As discussed in section~\ref{sec:inferencce}, we produce posterior predictive samples of the data, compute summary statistics and compare these with the observed ones. Table~\ref{table:stats} shows the sample mean, variance, interquartile range, and the 0.01 and 0.99 quantiles of the observed mortality counts along with corresponding estimates from the Poisson and Negative Binomial models. Clearly, the sample variance is underestimated by the Poisson model.

Furthermore, we compute a sequence of quantiles for the observations and the predictions to compare their overall distributions. Considering 200 equidistant quantiles (between 0 and 1, both included), Figure S2 indicates very good agreement between observations and (Negative Binomial) predictions, with the exception of the maximum observed count of 47, a rather large outlier. %Note that \texttt{mgcv} will produce the traditional residual QQ plot for each model, although for a large number of data points (3998 data points in our case) we find such plots difficult to interpret -- deviations from the diagonal can be masked when the data are big, in addition to repetitions in the observed counts.

\subsubsection{Temporal structure}
We also looked at whether the temporal structure (trend and auto-correlation) in the data was properly captured. To this end, we compare the sample auto-correlation function (ACF) of the observed counts, and compare this with the ACF of the predictions. Figure S3 (left panel) indicates that the observed autocorrelation values are consistently underestimated by the Negative Binomial model. This could be either due to trend not being captured or due to residual autocorrelation not being completely induced by the covariate. 

Exploratory plots (not shown) indicate a strong seasonal cycle, a weak day-of-the week cycle, a slight increasing trend with each year and fairly weak autocorrelation in the data. We thus expand the model by including: 1) a day-of-year, $doy(t)$, cyclic cubic spline (continuous between the 31/Dec and 01/Jan) with 48 coefficients; 2) an i.i.d. Gaussian random effect spline for day-of-week, ($dow(t)$); 3) a spline basis with 300 coefficients for the daily time step, $t$, that emulates a Gaussian Process with an power-exponential covariance function with power 0.05 in order to capture the temporal auto-correlation; and 4) a TPRS for year, $year(t)$, with 9 coefficients to capture inter-annual variability:
\begin{equation}
 \log(\mu_t) = \alpha + \sum_{l=0}^L h(l,x_{t-l}) + f_1(doy(t)) +  f_2(t) +f_3(dow(t))+ f_4(year(t)). \label{eq:ThesModel2.1}
\end{equation}
Fitting this using \texttt{mgcv} indicates that the edf was 6.4, 2.5,  71.7 and 1.2 for $f_1,\ldots,f_4$ respectively, confirming the exploratory analysis in that evidently there exists 1) seasonal variability; 2) temporal autocorrelation (71.7 out of 200 coefficients); 3) a weak day-of-week effect and 4) a positive and basically linear trend over the years. The corresponding ACF-checking plot confirms that autocorrelation is now well-captured (Figure S3).

\subsubsection{Model selection and expansion}
Other than using posterior simulations to compare between competing models, we can also use general measures of model comparison such as the Akaike Information Criterion (AIC): as estimate of out-of-sample prediction error. For instance, the AIC for the Poisson model was 23428, for the Negative Binomial it was 23405, while the lowest AIC of 23086 was for the Negative Binomial with temporal structures.

For model expansion, \texttt{mgcv} offers a rich variety of modelling options, such as interactions (as discussed in the next section), spatial structures (see supplementary code), random effects and a wide range of probability distributions. Some of these are illustrated in the following sections.

\subsection{Multiple lagged covariates}\label{sec:multiple_exposures}
The covariate (Tapp) used so far, is a function of temperature and relative humidity (RH), a ``pre-defined'' heat-stress metric. Here, we instead look at a daily mortality counts (2004--2019) from the island of Cyprus to explicitly investigate the joint effect of daily maximum temperature (Tmax) and RH (a percentage) along their lags. The counts are stratified by each of the 5 districts comprising Cyprus and we consider a Negative Binomial model with log mean:
\begin{equation} \label{eq:CyModel2}
\log(\mu_{t,s}) = \alpha + \sum_{l=0}^L h(l,x_{t-l,s},z_{t-l,s})  + \log\left(O(year(t),s)\right),
\end{equation}
where $x_{t,s}$ and $z_{t,s}$ are respectively Tmax and RH for district $s$ on day $t$, and $O(y(t),s)$ is the population offset in district $s$ for year $year(t)$. Function $h$ is the log-RR in terms of mortality rate per unit population and is constructed as a tensor of three TPRS with 10 knots each (with estimated edf of 33 from a total of 999 coefficients). The model assumes that the association of the covariates on mortality is the same for all districts (we return to this point later).

\begin{figure}[h]
\includegraphics[width=\textwidth]{"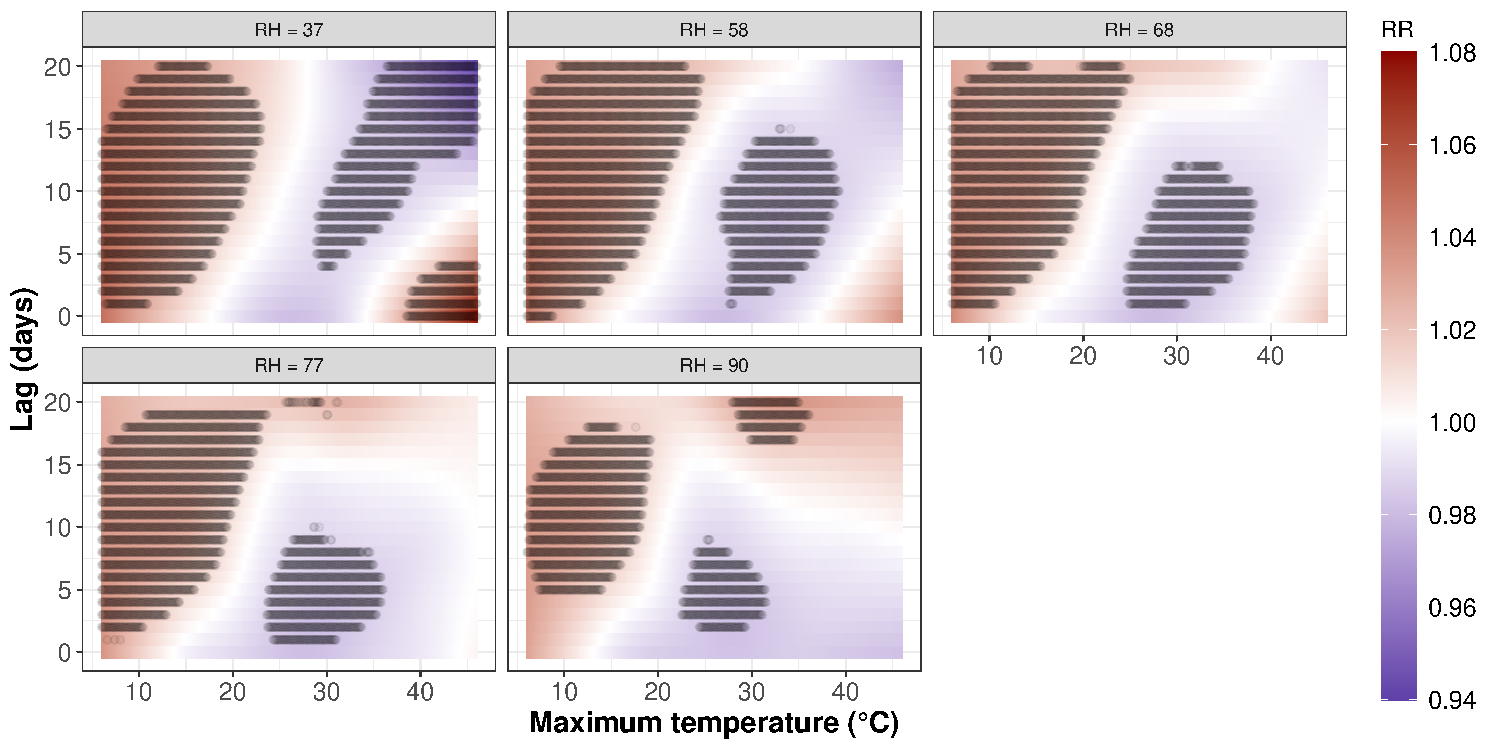"}
\caption{Relative risk estimates for specific values of relative humidity, specifically the 5\%, 25\%, 50\%, 75\% and 95\% sample quantiles.
\label{fig:figure3} }
\end{figure}

We plot the RR as a function of Tmax and the lag, for particular values of RH in Figure~\ref{fig:figure3}. The results do not support the hypothesis that hot-and-humid weather increases mortality risk~\cite{Cramer2022}, in fact the opposite is apparent. This may be due to adaptation (the majority of buildings in Cyprus are equipped with air-conditioning). In fact, the highest risk occurs at very high Tmax (>40$^\circ$) and very low RH (bottom right corner of top left panel). In other words hot and dry conditions are the worst (perhaps due to dehydration). This finding is however inline with other epidemiological studies of the health effects from humidity~\cite{Barreca2012,Zeng2017,Armstrong2019}. For Cyprus, the lowest RH values occur during warm months, which may explain the heat risk ``spike'' at low RH. Note also the elevated risk over low temperatures, across the range of RH values. For RH=90 (cold months) this occurs over medium to long lags and likely relates to increased spread of communicable diseases~\cite{Barreca2012}. There seems to be more risk over low temperatures which reflects the findings in the literature for other Mediterranean countries~\cite{Nastos2012TheEO, CONTI2005390}.

\begin{figure}[h]
\includegraphics[width=0.5\textwidth]{"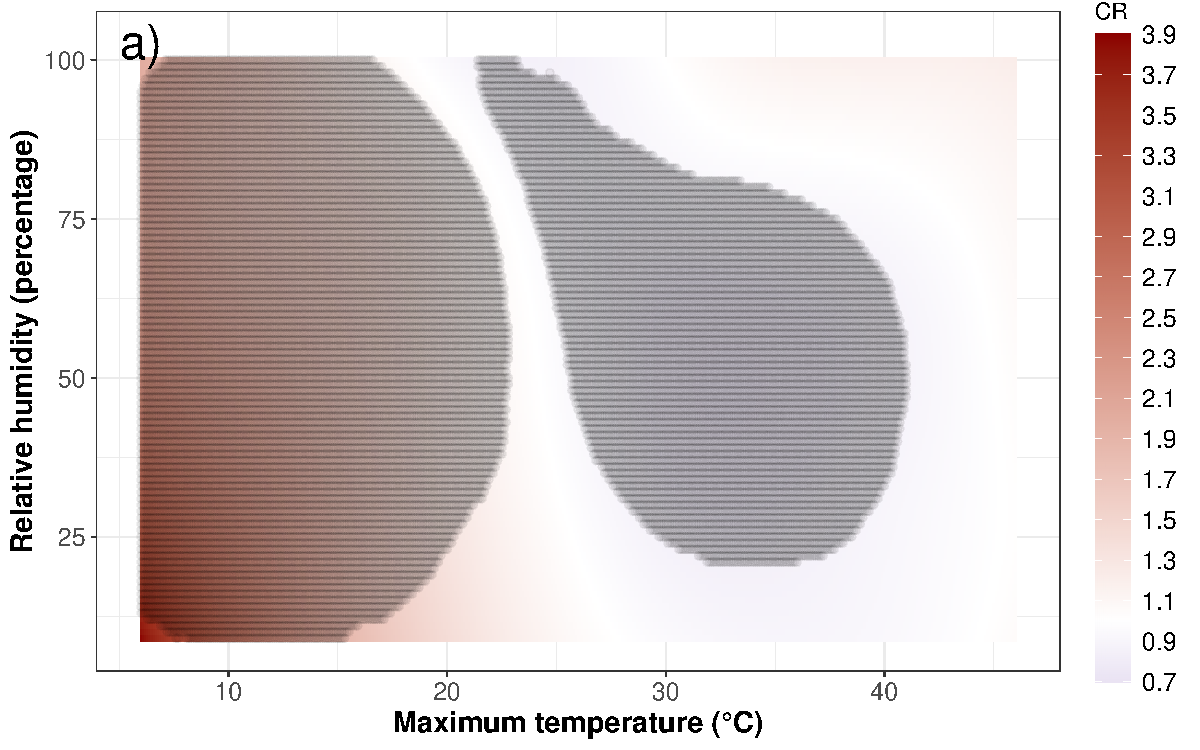"}
\includegraphics[width=0.5\textwidth]{"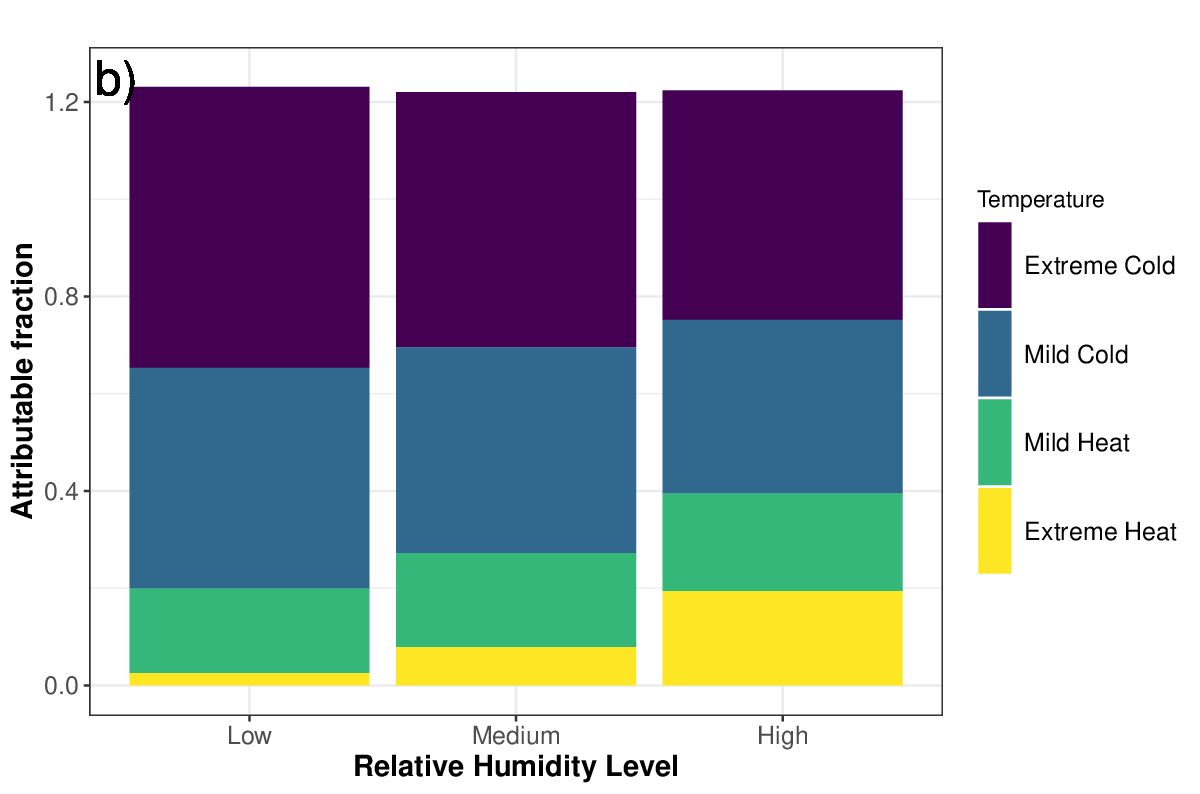"}
\caption{a) Cumulative risk surface depicting the interaction between Tmax and Relative Humidity. b) Attributable fraction for various regions of the RH-Tmax space. Tmax regions are defined as ``$<15.6$'', ``$[15.6,20.1]$'', ``$[31.2,36.0]$'' and ``$>36.0^\circ$C'' (based on the 5\%, 25\%, 75\% and 95\% sample quantiles). RH regions are ``$<58$'', ``$[58,77]$'' and ``$>77$\%'' based of the 25\% and 75\% sample quantiles. The bars are stacked and they do not sum to 100\%.
\label{fig:figure4} }
\end{figure}

The cumulative risk of Tmax and RH is shown in Figure~\ref{fig:figure4}a, where most of the risk accumulates in the low RH region, with a peak at low temperatures. There is also a mortality-risk ``sweet-spot'' for Tmax in [30,40] and RH in [25,75]. Figure~\ref{fig:figure4}b shows the AF for different Tmax-RH ``regions'', indicating increased mortality during extreme low temperatures, particularly for low RH. Hot-and-humid conditions do lead to elevated mortality, but the difference is small. The plots in Figures~\ref{fig:figure3} and \ref{fig:figure4} constitute novel insight into the joint effects of temperature and humidity (at least for Cyprus). Other covariates (e.g., air quality) can be added to investigate further interactions between covariates across lags.

\subsection{Hierarchical structures}
\begin{figure}[h]
\includegraphics[width=\textwidth]{"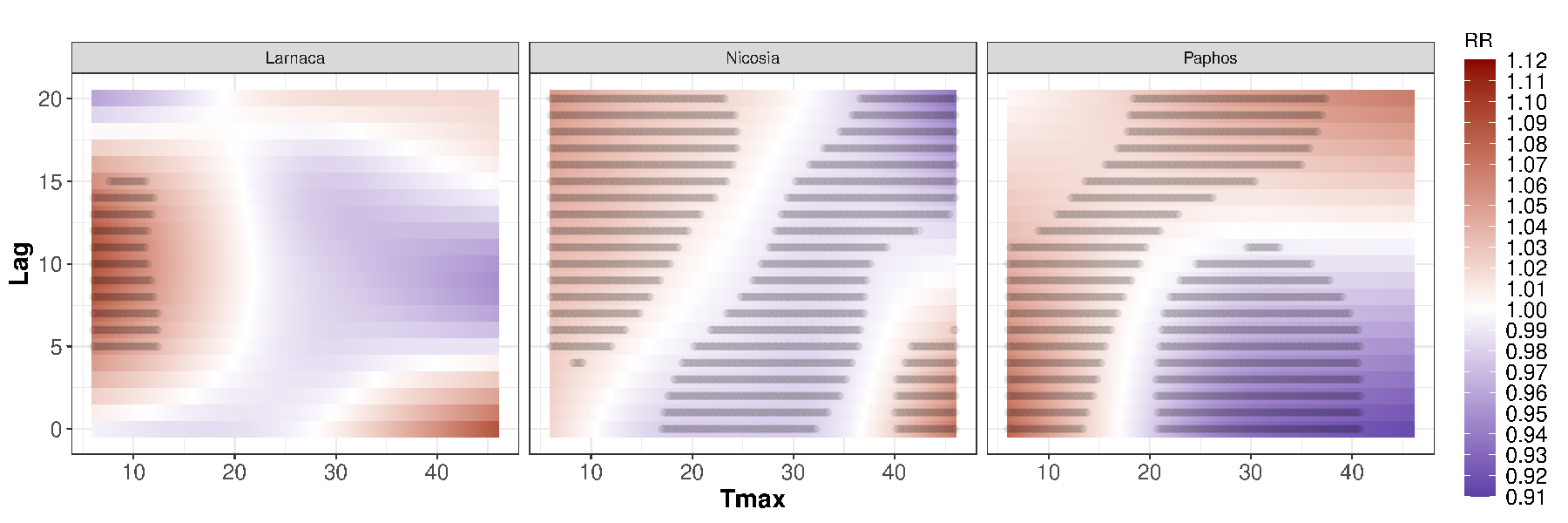"}
\caption{Relative risk estimates for 3 Cyprus districts, based on an individual model for each district.
\label{fig:figure5} }
\end{figure}

To illustrate interactions with categorical covariates using hierarchical structures (section~\ref{sec:hierarchies}), we consider estimation of district-specific risk for each of the 5 districts in Cyprus. Suppose we fit the following model to data from each of district:
\begin{eqnarray}
	y_t &\sim& NegBin(\mu_t,\theta)  \\
	\log(\mu_t) &=& \alpha + \sum_{l=0}^L h(l,x_{t-l}) + \log\left(O(year(t))\right)
\end{eqnarray}
where $x_t$ is maximum daily temperature and with 7 coefficients for each margin of the tensor. We would then notice unexpectedly large differences in the resulting RR surfaces. Figure~\ref{fig:figure5} shows the RR estimates for Nicosia (the capital, with population $\approx$350K in 2019), Larnaca (coastal, $\approx$150K) and Paphos (coastal, $\approx$95K). The differences are rather striking, given the proximity of these districts (Larnaca is about 50km, and Paphos is about 150km from Nicosia) and the size of the island (only 3.8\% the size of the UK). Even if we allow for the differences in population and weather, it is hard to scientifically justify why these surfaces are so different (with Paphos actually exhibiting RR $<1$ at extremely high temperatures). Close inspection of the data, reveals that the time series of counts for Paphos and Larnaca effectively comprise of ``zeros and ones", due to the low baseline mortality rate in these districts. We therefore hypothesize that the temperature signal is masked by this, and we must employ pooling across the districts to uncover the underlying risk.

To this end, we fit a hierarchical model where $s$ refers to the district:
\begin{eqnarray}
y_{t,s} &\sim& NegBin(\mu_{t,s},\theta)  \\
 \log(\mu_{t,s}) &=& \alpha + \alpha_s +\sum_{l=0}^L h(l,x_{t-l,s}) + \sum_{l=0}^L h_s(l,x_{t-l,s}).
\end{eqnarray}
We choose 7 knots for each margin, so that both the global term $h(l,x_{t-l,s})$ and the deviations $h_s(l,x_{t-l,s})$ have a-priori the same flexibility. Function $h_s(l,x_{t-l,s})$ is a tensor product interaction of two TPRS (for $x$ and $l$) and a ridge spline basis for $s$ that emulates an i.i.d. Gaussian random effect (which acts as a constraint across $s$ for each $h_s(l,x_{t-l,s})$. 

\begin{figure}[h]
\includegraphics[width=\textwidth]{"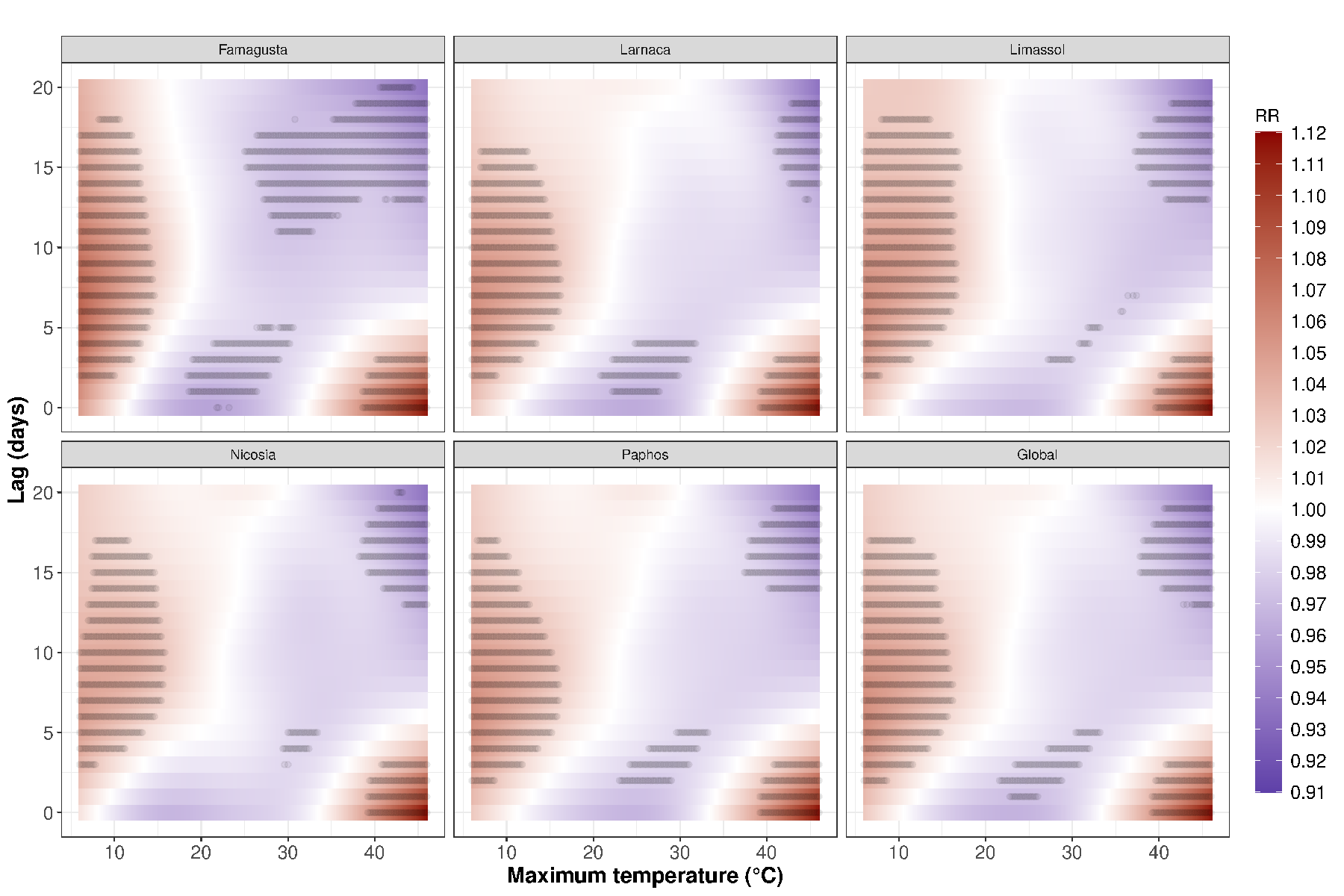"}
\caption{Pooled estimates of the RR for the Cyprus data. The bottom right plot is the estimate of the global term.
\label{fig:figure6} }
\end{figure}

Figure~\ref{fig:figure6} shows the resulting RR estimates for each district as well as the global term. With pooling, there is very little difference in the estimates across districts, as would be expected. Differences are just about noticeable for extreme low temperatures, with Famagusta having the highest risk. All districts exhibit a high risk ``spike'' at very high temperatures and low lags in addition to elevated risk at 3-15 lags for cold temperatures (disease transmission). The harvesting paradox (reduced risk at long lags for high temperatures) is more prominent for Famagusta in terms of significance. For more rigorous comparison between districts, we can compute the relative risk between any two, for instance $RR(l,x_{t}) = \exp\left\{ h_N(l,x_{t}) - h_F(l,x_{t}) \right\}$ where $N=$ Nicosia and $F=$ Famagusta (Figure S4). The AIC for a global model (same risk stricture for all districts) is larger by about 200 units relative to the hierarchical model indicating the although small, the differences across districts are substantial.

\subsubsection{Relation with existing approaches}
A current and widely-used approach to pooling DLNM estimates~\citep{Gasparrini2012,Gasparrini2013} is based on the concept of meta-analysis of spline coefficients. In summary, this involves the fitting of independent DLNMs (for each district, region, unit etc.) and then of another model to pool the estimated coefficients for each individual model (referred to as the two-stage approach). This pools coefficient estimates, and allows the inclusion of district-specific covariates that may explain the differences between the coefficients (and thus the mortality-covariate relationship) across the districts.

In principle, our approach can also be used to pool region-specific estimates (e.g., by simply taking the estimate of the global term in the hierarchical models) and to include district-specific covariates via tensor products, although this is beyond the scope of the current paper. Another difference is that the hierarchical approach produces estimates of the covariate-lag relative risk for each district, rather than the cumulative risk, enabling insight into how the whole covariate-lag risk profile varies with other variables (e.g., Figure~\ref{fig:figure3}) and Figure~\ref{fig:figure6}).

Lastly, the hierarchical approach involves a single model which, in conjunction with the Bayesian inference, provides full uncertainty quantification and thus interpretability of estimates via posterior simulation. This also allows for thorough model checking, which is important when choosing between competing models and for model expansion, as illustrated in section~\ref{sec:Thess}.

\section{Summary and conclusions}\label{sec:conclusion}
We presented the implementation of DLNMs as hierarchical GAMs, through the R package \texttt{mgcv}. The implicit penalisation in the GAMs ensures optimal estimates of the lagged effects (for a given maximum lag), which is important for robust estimation of risk. Through the flexibility offered by the \texttt{mgcv} package, we illustrated how the lagged effect from multiple covariates can be quantified (e.g., temperature and humidity effects on mortality). To the best of our knowledge, this is the first study to investigate the joint effects from temperature and relative humidity across temporal lags on human mortality. Moreover, we demonstrated that through the interpretation of penalised regression splines as random effects, one can perform approximate Bayesian inference, conditional on the penalty parameters and thus provide full uncertainty quantification, interpretable estimates and rigorous model checking. Lastly, we showed how hierarchical structures can be introduced, to pool the information across discrete units in the data (e.g., regions, age groups etc.) and provide robust estimates of the covariate-lag effect for each level of the discrete units, which constitutes a particular novelty of our proposed framework. 

The approach presented was inspired by and built upon many years of excellent work on using DLNMs as a useful data modelling tool across many areas (e.g., \cite{Zanobetti2000,Gasparini2010,Gasparrini2014,Obermeier2015}). The exposition of this framework provides a practically useful extension of the DLNM framework, particularly in the area of epidemiology, and provides a unifying approach that allows for penalisation of the smooth functions involved, pooling of the data, interactions between multiple covariates, Bayesian inference and, although not explicitly illustrated here, the ability for space-time structures in the covariate-response function. This latter point is left for future work, although in theory it is straightforward using the tensor product smooths we have demonstrated. An example of a smooth spatially varying covariate-lag function is given in the online supplementary material for simulated data.

As with any modelling framework, there are some caveats that need addressing in future work. On the more theoretical side, the Bayesian inference presented is conditional on both the penalty parameters and any hyperparameters of the conditional distribution (e.g., the ``size" parameter of the Negative Binomial). Although in our experience the uncertainty from these is relatively small (see full MCMC example provided in the Supplementary Material), this is still a potential issue when claiming to fully quantify uncertainty. On the practical side, computation efficiency or even feasibility can be an issue with the presented approach, particularly for a large number of coefficients in the splines. There are options in \texttt{mgcv} to deal with the ``many-coefficients'' and/or the ``many-covariates'' situation (e.g., function \texttt{bam} and parallel computing), demonstrated in the online supplementary material.

\newpage

%\backmatter

\section*{Acknowledgments}
Authors acknowledge support by the LIFE Programme of the European Union in the framework of the project LIFE21-GIE-EL-LIFE-SIRIUS/101074365. CS acknowledges support from the NIHR Health Protection Research Unit in Environmental Change and Health.

\subsection*{Author contributions}

TE: Conceptualization (lead), Formal Analysis (lead), Methodology (lead), Software, Writing – Original Draft Preparation. DP: Conceptualization, Data Curation, Formal Analysis, Validation, Visualization, Writing – Review \& Editing. AT: Conceptualization, Methodology, Review \& Editing. LD, OS, HS, RL, CS: Conceptualization, Formal Analysis, Software, Writing – Review \& Editing. MA: Conceptualization, Data Curation. JL: Supervision, Writing – Review \& Editing.

\subsection*{Financial disclosure}
TE and JL were funded by the European Union’s Horizon 2020 research and innovation programme under grant agreement No. 856612 and the Cyprus Government.\\
DP acknowledges the support provided by Greece and the European Union (European Social Fund-ESF) through the Operational Programme "Human Resources Development, Education and Lifelong Learning" in the context of the Act "Enhancing Human Resources Research Potential by undertaking
a Doctoral Research" Sub-action 2: "IKY Scholarship Programme for PhD candidates in the Greek Universities". \\

\subsection*{Conflict of interest}
The authors declare no potential conflict of interests.

\section*{Supporting information}
All work presented in this paper is completely reproducible via the online supplementary material, which includes data, code, instructions and supplementary figures. This can be accessed at Zenodo¬\cite{myData}. While the Cyprus and Thessaloniki health data cannot be made available, all models, estimates and model checking presented are reproduced for the open-access Chicago mortality data from the R package \texttt{dlnm}.

%The following supporting information is available as part of the online article:

%\noindent
%\textbf{Figure S1.}
%{500{\uns}hPa geopotential anomalies for GC2C calculated against the ERA Interim reanalysis. The period is 1989--2008.}

%\noindent
%\textbf{Figure S2.}
%{The SST anomalies for GC2C calculated against the observations (OIsst).}

%\appendix

%\begin{center}
%\begin{table}[b]%
%\centering
%\caption{This is an example of Appendix table showing food requirements of army, navy and airforce.\label{tab4}}%
%\begin{tabular*}{300pt}{@{\extracolsep\fill}lcc@{\extracolsep\fill}}%
%\toprule
%\textbf{col1 head} & \textbf{col2 head} & \textbf{col3 head} \\
%\midrule
%col1 text & col2 text & col3 text \\
%col1 text & col2 text & col3 text \\
%col1 text & col2 text & col3 text\\
%\bottomrule
%\end{tabular*}
%\end{table}
%\end{center}

%\nocite{*}% Show all bib entries - both cited and uncited; comment this line to view only cited bib entries;
%\clearpage
\bibliography{biblio}%

\clearpage
\begin{table}[h]
    \centering
    \begin{tabular}{lccc}
        Statistic &  Observations & Poisson  & Negative Binomial\\
         &  & predictions & predictions\\
        \hline
         Mean & 18.42 &18.43 [18.24, 18.61] &18.42 [18.23, 18.61]\\ 
         Variance & 22.39 &20.17 [19.24, 21.18]& 22.37 [21.31, 23.47]\\ 
         IQR & 6 & 6.07 [5.75, 7] & 6.39 [6, 7] \\ 
         1st quantile & 9 & 8.98 [8, 9]&8.58 [8, 9]\\ 
         99th quantile & 30 & 29.64 [29, 30]&30.35 [30, 31]\\ 
    \end{tabular}
    \caption{Summary statistics of the Thessaloniki mortality time series and corresponding estimates (posterior predictive means) from the Poisson and Negative Binomial models. 95\% prediction intervals are given in square brackets.}
    \label{table:stats}
\end{table}

%\section*{Author Biography}

%\begin{biography}{\includegraphics[width=66pt,height=86pt,draft]{empty}}{\textbf{Author Name.} This is sample author biography text this is sample author biography text this is sample author biography text this is sample author biography text this is sample author biography text this is sample author biography text this is sample author biography text this is sample author biography text this is sample author biography text this is sample author biography text this is sample author biography text this is sample author biography text this is sample author biography text this is sample author biography text this is sample author biography text this is sample author biography text this is sample author biography text this is sample author biography text this is sample author biography text this is sample author biography text this is sample author biography text.}
%\end{biography}

\end{document}